\documentclass[aps,prb,twocolumn,showpacs,preprintnumbers,amsmath,amssymb,superscriptaddress]{revtex4}

\usepackage{graphicx}
\usepackage{dcolumn}
\usepackage{bm}
\usepackage{figsize}
\def\mathbi#1{\textbf{\em #1}} 

\begin{document}


\title{Magnetotransport in Double Quantum Well with Inverted Energy Spectrum: HgTe/CdHgTe }

\author{M. V. Yakunin}
\email[]{yakunin@imp.uran.ru}
\affiliation{Institute of Metal Physics, 18 S. Kovalevskaya Str., Ekaterinburg 620990, Russia}
\affiliation{Ural Federal University, 19 Mira Str., Ekaterinburg 620002, Russia}
\author{A. V. Suslov}
\affiliation{NHMFL, FSU, 1800 East Paul Dirac Drive, Tallahassee, Florida 32310, USA}
\author{M.~R.~Popov}
\affiliation{Institute of Metal Physics, 18 S. Kovalevskaya Str., Ekaterinburg 620990, Russia}
\author{E.~G.~Novik}
\affiliation{ Physical institute, University of W\"{u}rzburg, D-97074 W\"{u}rzburg, Germany}
\author{S. A. Dvoretsky}
\affiliation{Institute of Semiconductor Physics, 13 Lavrentyev Ave., Novosibirsk 630090, Russia}
\affiliation{Tomsk State University, 36 Lenin Ave., Tomsk 634050, Russia}
\author{N. N. Mikhailov}
\affiliation{Institute of Semiconductor Physics, 13 Lavrentyev Ave., Novosibirsk 630090, Russia}
\affiliation{Novosibirsk State University, 2 Pirogov Str., Novosibirsk 630090, Russia}

\date{\today}

\begin{abstract}
We present the first experimental study of the double-quantum-well (DQW) system made of 2D layers with inverted energy band spectrum: HgTe. The magnetotransport reveals a considerably larger overlap of the conduction and valence subbands than in known HgTe single quantum wells (QW), which may be regulated in some extent by an applied gate voltage $V_g$. This large overlap manifests itself in a much higher critical field $B_c$ separating the range above it where the quantum peculiarities shift linearly with $V_g$ and the range below with a complicated behavior. In the latter case the $N$-shaped and double-$N$-shaped structures in the Hall magnetoresistance $\rho_{xy}(B)$ are observed with their scale in field pronouncedly enlarged as compared to the pictures observed in an analogous single QW. The coexisting electrons and holes were found in the whole investigated range of positive and negative $V_g$ as revealed (i) from fits to the low-field $N$-shaped $\rho_{xy}(B)$, (ii) from the Fourier analysis of oscillations in $\rho_{xx}(B)$ and (iii) from a specific behavior of the quantum Hall effect. A peculiar feature here is that the found electron density $n$ remains almost constant in the whole range of investigated $V_g$ while the hole density $p$ drops down from the value a factor of 6 larger than $n$ at extreme negative $V_g$ to almost zero at extreme positive $V_g$ passing through the charge neutrality point. We show that this difference between $n$ and $p$ stems from an order of magnitude larger density of states for holes in the lateral valence band maxima than for electrons in the conduction band minimum. We interpret the observed reentrant sign-alternating  $\rho_{xy}(B)$ between electronic and hole conductivities and its zero resistivity state in the quantum Hall range of fields on the basis of a calculated picture of magnetic levels in a DQW. This behavior is due to (i) an oscillating of the valence subband top versus field in the overall picture of its magnetic levels, (ii) a reduced gap between the lowest electron and the highest hole magnetic levels where the electron- and hole-type localized states are superposed, and (iii) a possible formation of the interlayer electron-hole excitons.
\end{abstract}

\maketitle

\section{\label{sec:level1}Introduction}

\begin{figure*}[t]
\includegraphics[width=\textwidth]{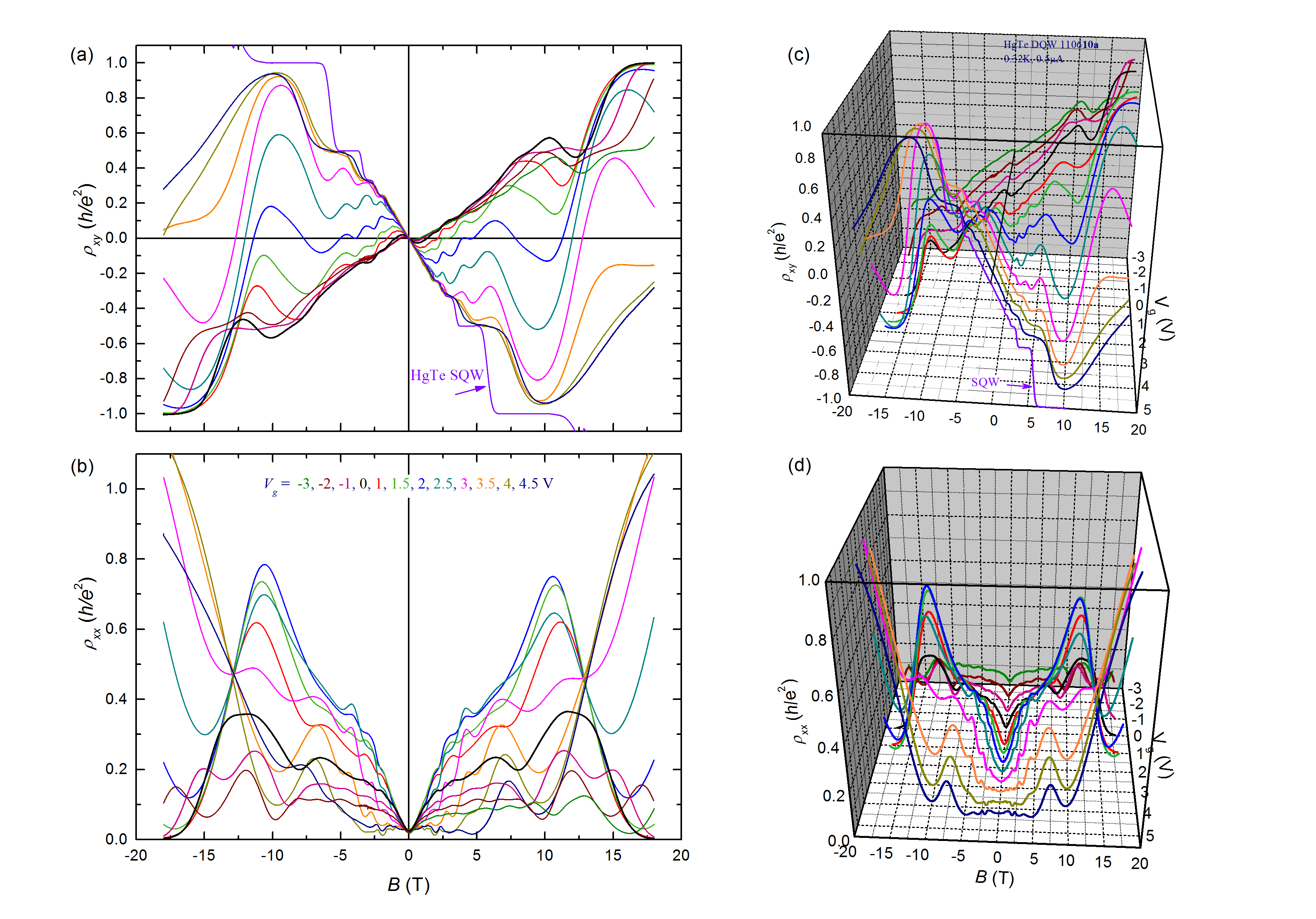}
\caption{\label{fig:VgXX-XY}  (color online). (a) The Hall $\rho_{xy}(B)$ and (b) diagonal $\rho_{xx}(B)$ magnetoresistances of the DQW Sample~1 in a series of gate voltages $V_g=(-3\div+4.5)$~V. The colors of digits for the $V_g$ values in the figure correspond to the colors of relevant curves, and this color system is retained in other corresponding figures. (c) and (d) -- the same in the bulk graph presentations. $\rho_{xy}(B)$ at large positive $V_g$ are compared with the QH curve for an identical Single QW.}
\end{figure*}

Spatially quantized energy spectrum of the HgTe layer wider than $d_w=d_c\approx 6.3$~nm is markedly specific due to its inversion character.\cite{Konig} In this case the conduction and valence subbands are close in energy due to their quasidegeneracy and for $d_w\gtrsim 8.3$~nm both are built of the $\Gamma_8$-symmetry wave functions contrary to the traditional semiconductor structures with their finite intrinsic gap and a conduction subband of the $\Gamma_6$ character. Moreover, the conduction subband may overlap here with the lateral maxima of the valence subband\cite{Kvon JETPLett2008} leading to formation of a semimetal with coexisting electrons and holes. The overlap occurs in a relatively wide HgTe quantum well (QW) depending on the heterostructure characteristics, of which probably the main is the built-in deformation.\cite{Kvon PRB2011,Olshanetsky SSC2012} Diverse new physics arises in such a semimetal.  For example, correlations between the electron and hole subsystems may cause a transition into a state of exciton insulator\cite{OlshanetskyJETPL2013} while the enhanced electron-hole friction in this situation results in modified temperature dependencies of transport coefficients.\cite{EntinJETP2013}  The picture of magnetic levels in this semimetal is very intricate as it consists of two overlapped fan-charts  oppositely directed in energy. Additional complications appear here due to existence of specific zero-mode levels  of a strongly mixed electron-hole nature in each subband and due to a nonmonotonous behavior of the valence subband levels around the energy of lateral maxima. A gap is opened between the lowest electron magnetic level and the highest hole levels at some critical magnetic field $B_c$ thus creating a possibility to realize a unique zero filling factor state when the Fermi level gets into this gap.\cite{Gusev-2010,Raichev-2012} This state has also  been revealed  in graphene and has been actively debated.\cite{Zhang2006} The specific shape of the valence subband with a central minimum and lateral maxima causes an oscillating-like behavior of the subband top profile vs. quantizing magnetic field. These oscillations may be felt in the quantum magnetotransport once their scale is comparable with the typical distances between the valence subband levels when the quantum features connected with these levels are resolved experimentally. Physics get especially unusual here because of a possible competition between the states of a heavy electron character in the central valence subband minimum and the hole states around its lateral maxima and at larger wave vector values.

We present our investigations of a technique to change and regulate \textit{in situ} the subband overlap using a system of two relatively wide HgTe quasi-2D layers separated by a thin CdHgTe barrier, i.e., a HgTe double quantum well (DQW) with inverted energy spectrum.  Application of a gate voltage $V_g$ causes a shift of the whole energy spectrum picture in the upper HgTe layer (which is closer to the gate) with respect to a similar energy spectrum of the lower HgTe layer, since the free charges in the upper layer screen the lower layer so that its energy is almost insensitive to $V_g$.\cite{DQWlayershifts} It results in a change of the overlap between the conduction ($c$) subband in one layer and the valence ($v$) subband in the other. The HgTe/CdHgTe system is specific in this sense differing from a traditional DQW, e.g., the GaAs/AlGaAs system, since for normal gaped semiconductors the interlayer overlap occurs only for the same kind of bands: $e-e$ or $h-h$. In a DQW of traditional semiconductors the exhaustion of the upper layer may be achieved, after what the charge density in the lower layer starts to vary with $V_g$.\cite{DQWlayershifts} In contrast, the exhaustion of the upper layer would never come in a HgTe layer, only the kind of free charges will be changed here when the Fermi level $E_F$ moves between the touched or overlapped $c$- and $v$-bands.

\begin{figure*}[t]
\includegraphics[width=\textwidth]{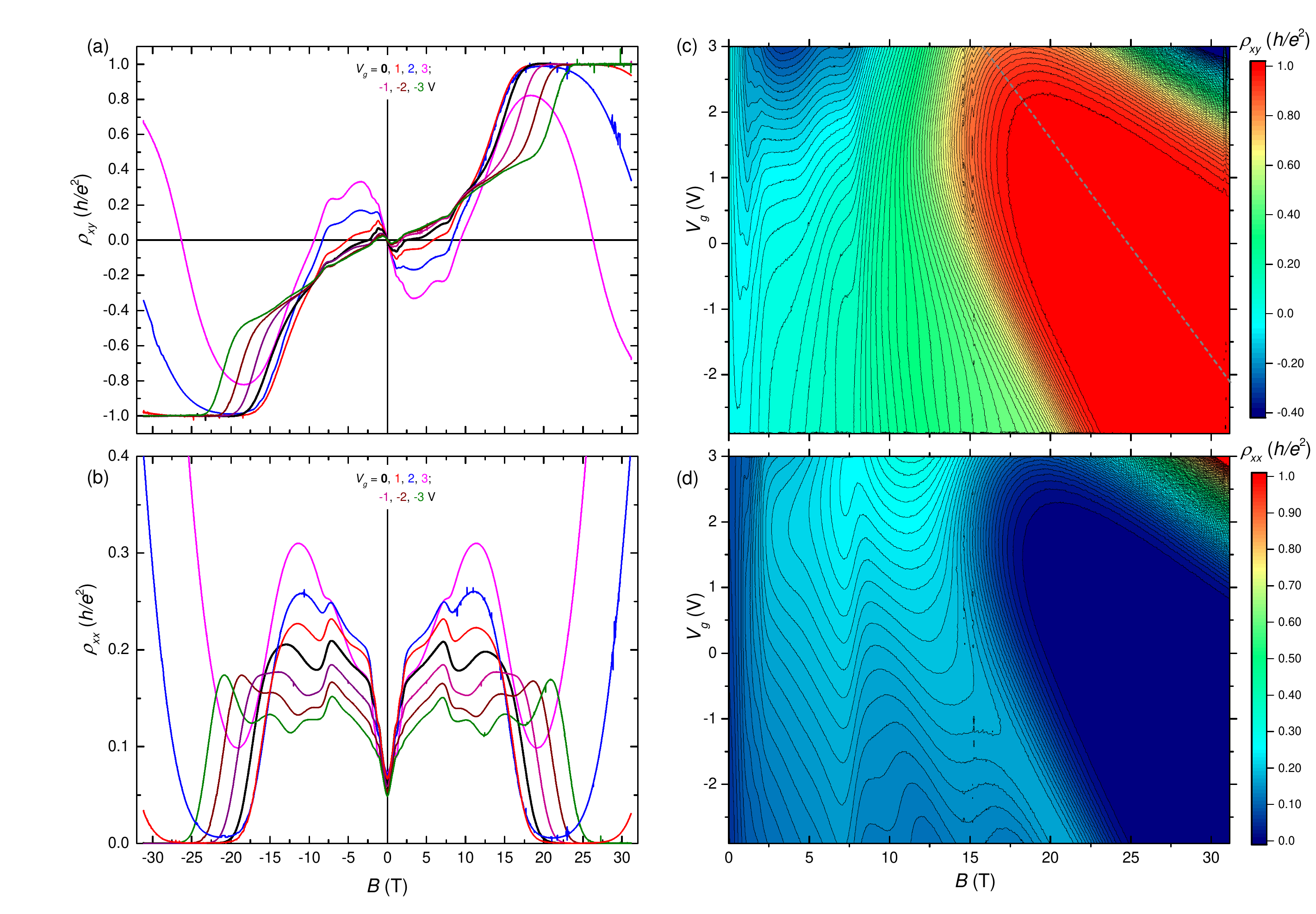}
\caption{\label{fig:SawS}  (color online). (a,c) The Hall $\rho_{xy}(B)$ and (b,d) diagonal $\rho_{xx}(B)$ magnetoresistances of the DQW Sample~2. Figures (a) and (b) are for the series of fixed values of $V_g=-3\div+3$~V  while (c) and (d) are obtained by the continuous coordinated sweeps of $V_g$ and $B$ and presented in the color scale map view. The inclined dashed line in (c) marks the middle of the $i=+1$ QH plateau as it linearly shifts in $B$ with $V_g$.} 
\end{figure*}

An exclusion is the InAs/GaSb system where two normal spectra are intrinsically so much shifted in energy that the InAs $c$-band falls below the GaSb $v$-band top.\cite{Nicholas-2000,Suzuki-2003} However, a substantial interlayer hybridization gap appears in this system, which considerably modifies the collective energy spectrum. Also, a balanced state with equal electron and hole densities (a charge neutrality point, CNP), which is one of the most remarkable moments in a system with overlapped $c$- and $v$-bands, has not been reached in InAs/GaSb system so far.\cite{Suzuki-2004} 

Although the overlap of $c$ and $v$ bands occurs in our case  \textit{between} the layers, it affects interlayer electron--hole ($e-h$) interactions in the DQW with a thin enough barrier thus predetermining magnetotransport in the whole system.  Variation of $V_g$ in a DQW provides some new opportunities to scan the complicated structure of the valence subband top by the Fermi level. Information about the valence subband top in the case of overlapped bands is especially actual since some controversy about its structure has been revealed recently.\cite{Minkov-2013} 

Other important aspects of the HgTe DQW are caused by the fact that the magnetic levels of different momentum projections in the $\Gamma_8$ conduction subband, which are analogs of the spin sublevels in a $\Gamma_6$ band, are easily resolvable here indicating a large effective $g$-factor\cite{YaPE2010} and leading to well pronounced effects of spin polarization,\cite{YaPRB2012} i.e., to a good realization of the spin degree of freedom. It invokes a question: how would manifest itself such an unusual combination of the HgTe layer characteristics in the collective electronic properties of DQW? Existence of an additional pseudospin degree of freedom connected with the possibility for an electron to reside in one of the two interconnected layers is known to result in a formation of new correlated states of the electronic system\cite{Girvin} or in a improved stabilization of the correlated states known in a single QW as a quantum Wigner solid.\cite{Manoharan-96} Also shown was\cite{Giudici-2008} that a well pronounced spin degree of freedom plays a positive role in formation of correlated phases. Would the $p$-orbitals of the HgTe $\Gamma_8$ conduction and valence subbands introduce some novel features in the possible correlation effects? Would the addition of one more degree of freedom -- a well pronounced effective spin degree -- manifest itself here? The interest to these subjects is still enhanced  by the prediction\cite{Michetti-Novik2012-2014} that a special kind of topological insulator may be created in the HgTe DQW with a perspective of its utilization in spintronics devices. We are doing the experimental attempts to answer some of these questions as presented in this work.

\section{\label{sec:level2}Experimental details}

The samples are grown on a  (013)-oriented GaAs substrate with their active part consisting of two HgTe layers, 20~nm wide, separated by a $6\div10$~nm Cd$_{0.65}$Hg$_{0.35}$Te barrier. The DQW is symmetrically doped with In in the adjacent outer barriers at the distances of 10 nm from the outer interfaces. 
Here we outline the results for two samples, which are representatives of two investigated groups differed mainly by the hole densities $p$. The first one (Sample~1) having a moderate $p=6.5\times10^{15}$ m$^{-2}$ at $V_g=0$ and the other (Sample~2) with a larger amount of Hg vacancies resulted in estimated larger $p=8\times10^{15}$ m$^{-2}$.
The samples are shaped in the double Hall bridge and covered with a gate above the silicon insulating layer, as in Ref.~\onlinecite{Raichev-2012}. The results on the DQW are compared with those obtained on a single QW with identical parameters grown in the same technological cycle (20~nm wide, the same symmetrically doped). Diagonal and Hall magnetoresistances (MR), $\rho_{xx}(B)$ and $\rho_{xy}(B)$, were measured simultaneously by the direct current reversal technique using an improved version of original experimental setup \cite{Suslov} on a dc current of $0.25\div2$~$\mu$A under magnetic fields up to 31~T at a temperature of 0.31~K at a fixed or sweeping $V_g$: Figs.~\ref{fig:VgXX-XY},~\ref{fig:SawS}. In the latter case a continuous sequence of bipolar triangular or sawtooth pulses  was applied to the gate with amplitudes corresponding to the predetermined range of $V_g$ at a frequency an order of magnitude smaller than the measuring rate at a sufficiently low field sweep. Based on these data, $\rho_{xy}(B,V_g)$ and $\rho_{xx}(B,V_g)$ were built as continuous functions of two variables: see Figs.~\ref{fig:SawS}(c,d).

The curves for $\rho_{xy}(B)$ and $\rho_{xx}(B)$  at weak fields for negative and small positive voltages $-3$~V~$\leqslant V_g\lesssim$~+1~V have a structure relevant to the classical model of two kinds of free charges -- a small quantity of high mobility electrons and a large quantity of low mobility holes, i.e., the \textit{N}-shaped  $\rho_{xy}(B)$ and a concomitant parabolic $\rho_{xx}(B)$ (Figs.~\ref{fig:VgXX-XY},~\ref{fig:SawS}), as it was first found in Ref.~\onlinecite{Kvon JETPLett2008} for this heterosystem. MR oscillations are superimposed on this classical background. Within this range of $V_g$, the positive  $\rho_{xy}(B)$ reaches the quantum Hall (QH) plateau at $h/e^2$ for the hole component of conductivity in the highest fields simultaneously with  $\rho_{xx}(B)$ dropping to zero. Also some feature exists in $\rho_{xy}(B)$ at  $h/2e^2$ representing a distorted $i=+2$ QH state of holes.  

\begin{figure}[b]
\includegraphics[width=90mm]{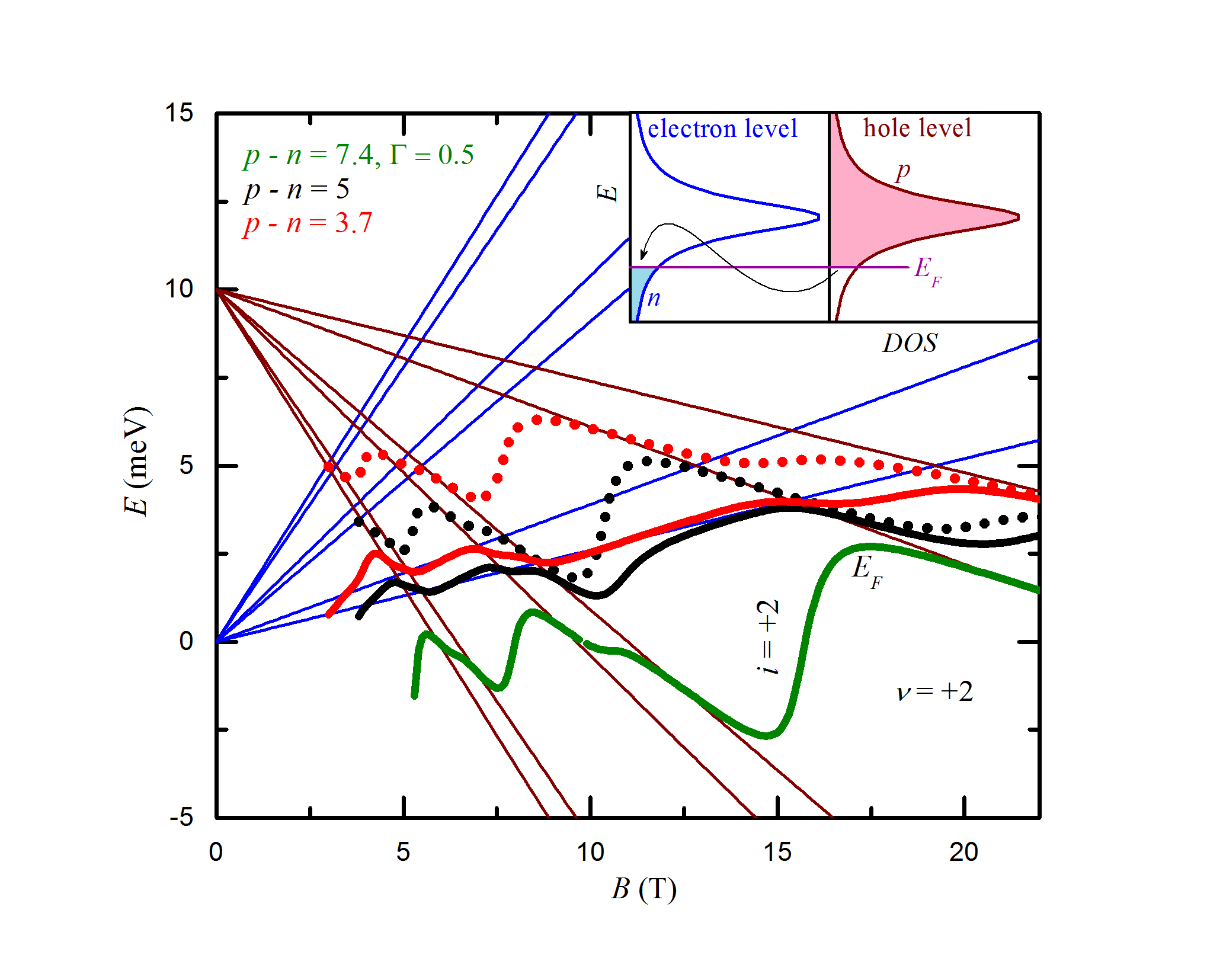}
\caption{\label{fig:Model_v-c}  (color online). Calculated evolution of the Fermi level within the modeled overlapped fan charts of the hole and electron magnetic levels with decreasing hole density $(p-n)$ (solid lines) compared to the corresponding cases without electron levels: $n=0$ (dotted lines). The densities are indicated in the picture in $10^{15}$ m$^{-2}$. Lorentzian shape of the density of states in the level of the width $\Gamma=0.5$~meV is supposed. Inset: Populations of the electron and hole levels by the free electrons and holes correspondingly. The arrow illustrates the transition of electrons from the lower part of the hole level into the electron level as the Fermi level is moving up across both of them.}
\end{figure}

There is a striking difference in the evolution of $\rho_{xy}(B)$ with $V_g$ within the QH range of fields for negative and small positive $V_g$ on one side and for the large $V_g$ on the other. The QH features in the hole conductivity component move in field linearly with $V_g$ in the former case as it should be for a regular picture of magnetic levels and a constant capacitance between the gate and the two-dimensional conducting system. To illustrate this linear dependence the dashed inclined line is drawn through the middle of the $i=+1$ QH plateau area for Sample 2 in Fig.~\ref{fig:SawS}(c). Also such a linear shift is seen for the $i=+1$ to $i=+2$ plateau-plateau transition (PPT) in $\rho_{xy}(B)$ for both samples in Figs.~\ref{fig:VgXX-XY}(a) and ~\ref{fig:SawS}(a). But more complicated behavior shows up for the further increase of $V_g$. Instead of further shrinkage of the field scale as is expected in a simple case for the decreasing $p$, we observe that the mentioned PPT stops moving to weaker fields at $\sim15$~T and $\rho_{xy}(B)$ \textit{gradually} sinks to negative values, passing through the states where several sign reversal  points of  $\rho_{xy}(B)$ exist (Fig.~\ref{fig:VgXX-XY}). Simultaneously the $\rho_{xx}(B)$ minimum at the highest fields moves up from zero. This behavior is quite different from that observed in a single QW where the transition from a positive to negative $\rho_{xy}(B)$ of an almost undistorted QH shape occurs rather sharply (see, e.g., Ref.~\onlinecite{Konig-2007Science} for a narrow QW and Ref.~\onlinecite{Minkov-2013} for a 20~nm wide well). 

\section{\label{sec:level3}Discussion of the results}

\begin{figure}[b]
\includegraphics[width=\columnwidth]{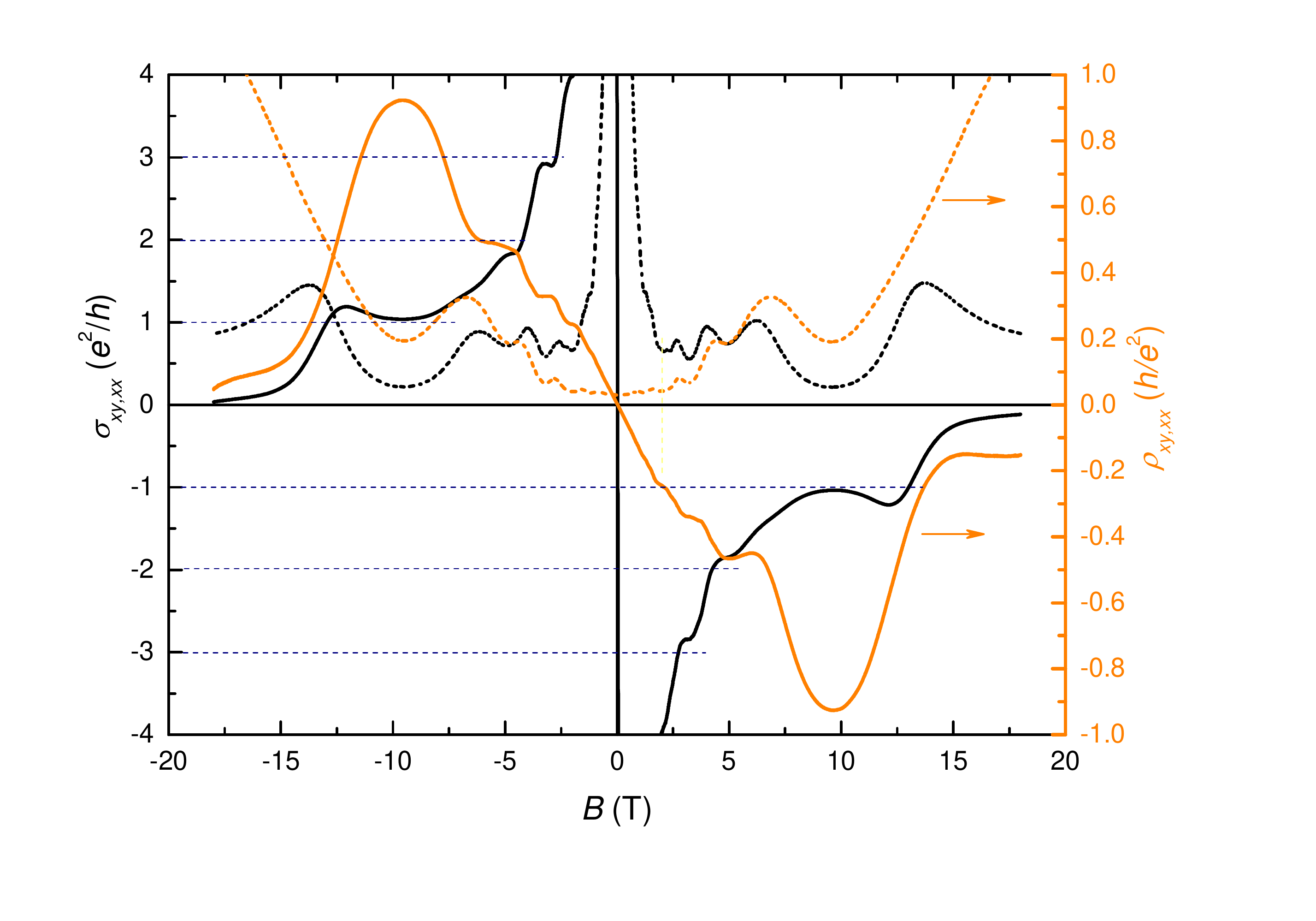}
\caption{\label{fig:SXX-XY}  (color online). MR and conductivities of Sample~1 for $V_g=+3.5$~V. Notice a drop of $\sigma_{xy}(B)$ to zero at the highest fields with concomitant minimum of $\sigma_{xx}(B)$.}
\end{figure}

The described behavior may be explained on a simplified scheme of magnetic levels as resulted from a superposition of the fan chart of electron levels upon the hole levels:  Fig.~\ref{fig:Model_v-c}. Indeed, when at large enough $p$ (corresponding to our case of negative $V_g$) $F_F$ is sufficiently deep in the valence subband, the superposed electron levels are empty, as if they are absent at all, and we have only a regular fan chart of the populated hole magnetic levels. In this case QH features would move linearly with decreasing $p$ and consequently -- with increasing $V_g$. We modeled a finite width $\Gamma$ of the levels by introducing a tentative Lorentzian density of states. Using Gaussian or semi-elliptic shape would not radically change the results. A QH feature in this picture corresponds to the range of fields where $E_F$ is sufficiently distant from both the neighboring levels when it passes through the gap between them. Introducing the finite level width makes it possible to display the finite sises of the QH features, but the main thing is that these calculations allow to represent automatically the behavior of $E_F$ when the electron levels come into play with the hole level. In the picture of overlapping hole and electron levels, a hole level is populated by free holes above $E_F$ simultaneously with free electrons occupying the electron level below $E_F$ (Fig.~\ref{fig:Model_v-c}, inset).

When, on decreasing $p$, the $E_F$ comes into touch with an electron level and this level starts to populate with free electrons, the corresponding QH feature stops moving towards lower fields. To elucidate this behavior the $E_F$ curves calculated for the full set of levels (solid lines in Fig.~\ref{fig:Model_v-c}) are compared with those calculated for the same values of $p-n$ but for the case as if there are no electron levels:  $n=0$, dotted lines. It is seen that population of electron levels considerably modifies the positions of QH features stopping them in a vicinity of the point where the corresponding hole level crosses with an electron level. When $E_F$, on moving along a hole level with decreased field, reaches the crossing electron level, then a part of electrons occupying the hole level below $E_F$ starts to move into the electron level (Fig.~\ref{fig:Model_v-c}, inset) so that $n$ appears and increases. That is why, at fixed $p-n$, $p$ increases and the corresponding QH feature within a set of hole levels appears at the fields higher than the position where it would be without the electron level. This process hampers the further shift of QH features with increasing $V_g$ to lower fields within a set of hole levels. Thus the experimental field where the QH features stop moving with $V_g$ indicates the approximate position of some crossing point in the picture of levels. Another important consequence appearing when $E_F$ approaches the electron level is an increased contribution of electron states into the magnetotransport that would manifest in a development of $\rho_{xy}(B)$ to negative values.

The reentrant sign-alternating behavior of $\rho_{xy}(B)$ has been revealed in a single HgTe QW in Ref.~\onlinecite{Raichev-2012} at much lower fields of several Tesla, and a value of the band overlap in their sample was estimated as of the order of 1~meV. Much higher fields of the $\rho_{xy}(B)$ sign reversal in our DQW structures indicate much larger overlaps and much higher values of characteristic fields for the level crossings.

For the largest  $V_g$, $\rho_{xy}(B)$ of Sample~1 becomes fully negative and reveals a series of QH plateaus  at intermediate fields due to electron conductivity component. However  $\rho_{xy}(B)$ \textit{turns back towards zero} at the highest fields [Fig.~1(a,c)] reaching a plateau-like feature close to zero. To emphasize the difference of this anomalous behavior of  $\rho_{xy}(B)$ in the DQW from the traditional one, the trace for a single $n$-type HgTe QW (measured without a gate) is added to Figs.~\ref{fig:VgXX-XY}(a,c). Concomitantly with the drop of $\rho_{xy}(B)$ in a DQW, a sharp growth appears in $\rho_{xx}(B)$ contrary to the behavior in the traditional QH effect (QHE). This quasi-zero plateau in $\rho_{xy}(B)$ is well reproduced in the conductivity (Fig.~\ref{fig:SXX-XY}) and may be an indication of a zero-filling-factor state.\cite{Gusev-2010} We don't see the electron QHE and its further development in Sample~2 because of a smaller achievable $V_g=+3$~V due to a less reliable insulator under the gate in it. Nevertheless the reentrant sign change of $\rho_{xy}(B)$ combined with the sharp growth of $\rho_{xx}(B)$ are well seen in Fig.~\ref{fig:SawS} at maximal fields and the maximum positive $V_g$. 

In the range of the largest  $V_g$, when a considerable electron part of $\rho_{xy}(B)$ is formed up to intermediate fields, remarkable is that the weak field slope of this curve does not change with $V_g$, contrary to the data for single QWs\cite{Konig-2007Science,Minkov-2013} where the increase of free charge density with $V_g$ manifests in a monotonic decrease of its slope, i.e., in a monotonic decrease of the Hall voltage.

\subsection{\label{sec:level3a}Classical treatment}

To estimate the parameters of free charges in our structures we first used a simple classical model for magnetotransport with two kinds of carriers (see, e.g., Ref.~\onlinecite{Seeger}) -- electrons with density $n$ and mobility $\mu_e$ and holes with density $p$ and mobility $\mu_h$:

\begin{subequations}
\label{eq:2r}
\begin{equation}
\rho_{xy}=\frac{B}{|e|}\frac{(p-n)\mu_h^2\mu_e^2B^2+(p\mu_h^2-n\mu_e^2)}{(p-n)^2\mu_h^2\mu_e^2B^2+(p\mu_h+n\mu_e)^2},
\label{eq:2rxy}
\end{equation}
\begin{equation}
\rho_{xx}=\frac{1}{|e|}\frac{(p\mu_e+n\mu_h)\mu_h\mu_eB^2+(p\mu_h+n\mu_e)}{(p-n)^2\mu_h^2\mu_e^2B^2+(p\mu_h+n\mu_e)^2}.
\label{eq:2rxx}
\end{equation}
\end{subequations}

To extract these four parameters the following four quantitative experimental features in $\rho_{xy}(B)$ and $\rho_{xx}(B)$ were used: (i) a zero field resistivity $\rho_0$; (ii) a slope of the Hall resistivity at high fields $R_H^{\infty}$ where it should approach the value $1/e(p-n)$ in the classical model; (iii) a slope of the Hall resistivity around zero $R_H^0$ and (iv) a field of the first inversion $B_0$. Then a system of four equations for four unknowns reads:

\begin{figure}[b]
\includegraphics[width=\columnwidth]{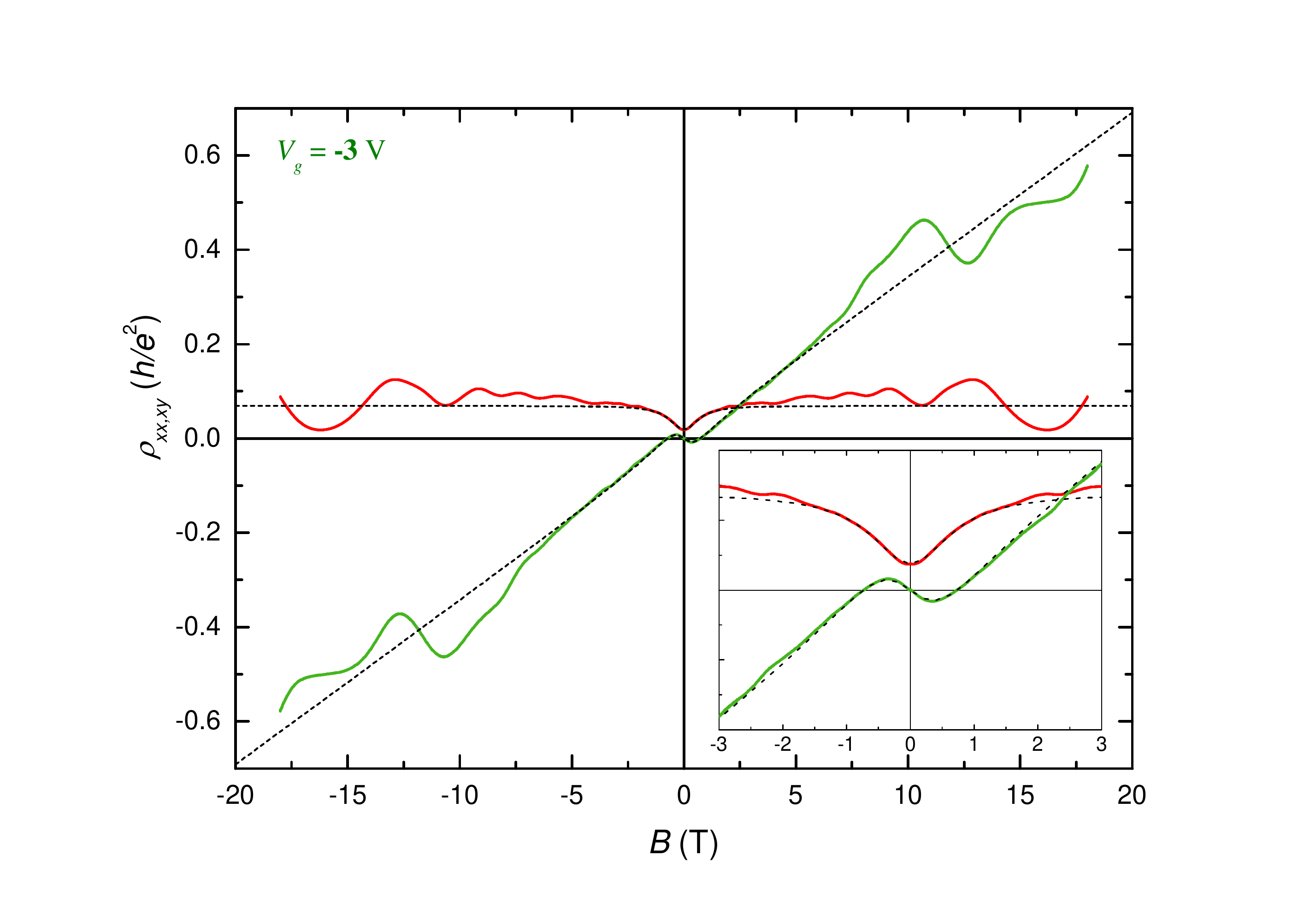}
\caption{\label{fig:G111}  (color online). Two-carrier fit to MR of Sample 1 for $V_g=-3$~V with an enlargement in the inset.}
\end{figure}

\begin{subequations}
\label{eq:4eqtns}
\begin{equation}
p-n=\frac{1}{eR_H^\infty},
\label{eq:np}
\end{equation}
\begin{equation}
n\mu_e+p\mu_h=\frac{1}{e\rho_0},
\label{eq:nmue-pmuh}
\end{equation}
\begin{equation}
\mu_e\mu_h=\frac{\sqrt{-R_H^0R_H^\infty}}{\rho_0B_0},
\label{eq:mue-muh}
\end{equation}
\begin{equation}
n\mu_e^2-p\mu_h^2=-\frac{R_H^0}{e\rho_0^2},
\label{eq:nmue2-pmuh2}
\end{equation}
\end{subequations}
which was solved numerically.

\begin{figure}[b]
\includegraphics[width=125mm]{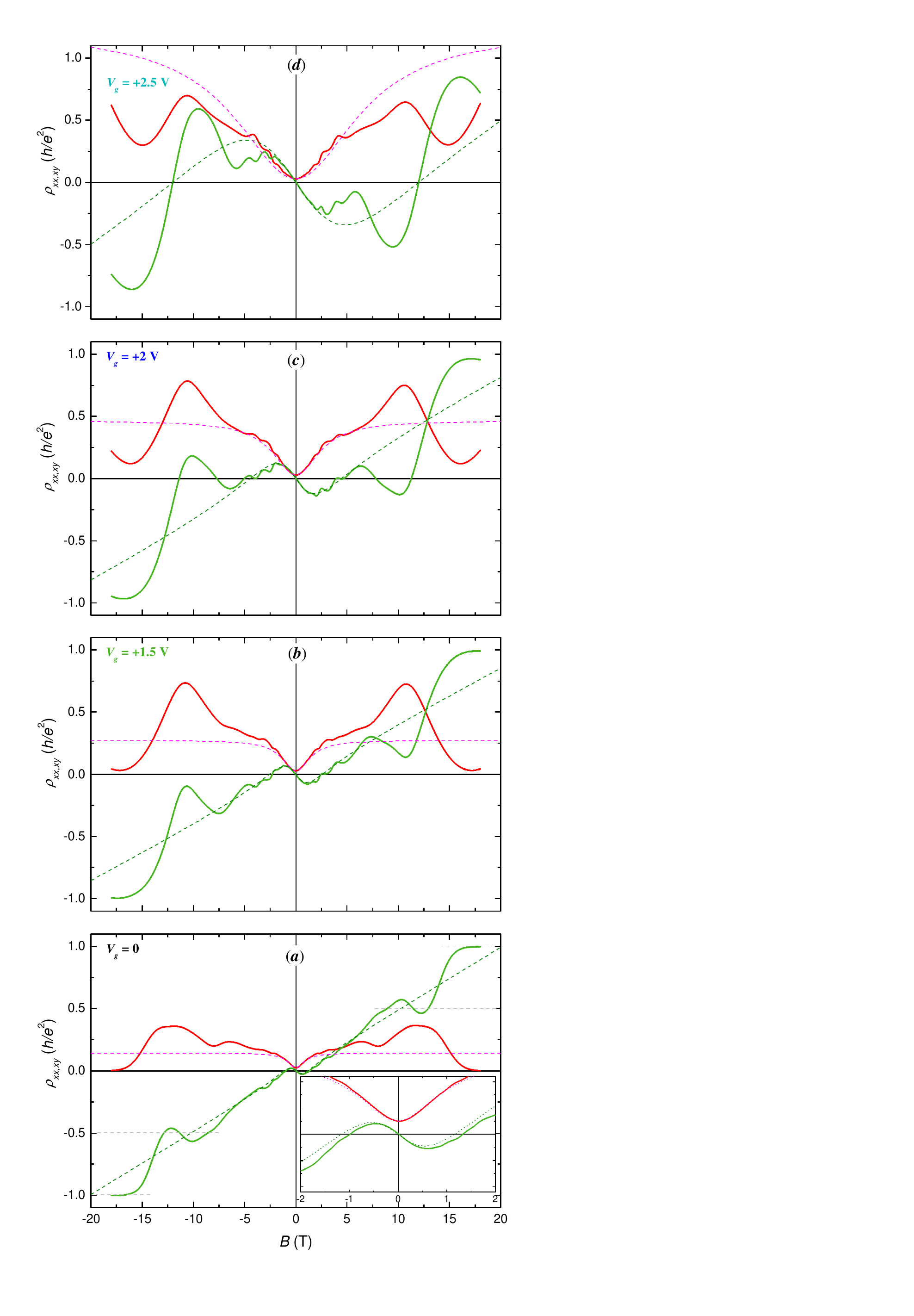}
\caption{\label{fig:e-h_Sum}  (color online). Two-carrier fits to MR of Sample 1 for: (a) $V_g=0$ with an enlargement in the inset; (b) $V_g=+1.5$~V; (c) $V_g=+2$~V and (d) $V_g=+2.5$~V. Notice the ambiguities in 2-carrier fits for the large positive $V_g$ values.}
\end{figure}

The classical description should be valid at weak fields while deviations from it may appear at higher fields caused by unconsidered complications due to quantization. This fit indeed works in our samples at the negative $V_g$. Thus, for Sample 1, the weak field fit is excellent for $V_g=-3$~V (Fig.~\ref{fig:G111})  with $n=1.89 \times 10^{15}$~m$^{-2}$, $\mu_e=3.8$~m$^2$/V$\cdot$s, $p=8.87 \times 10^{15}$~m$^{-2}$, $\mu_h=0.66$~m$^2$/V$\cdot$s, while at higher fields the Shubnikov - de Haas oscillations (SHO) are superposed on the classical horizontal line for $\rho_{xx}(B)$ and a deviation from the inclined line for $\rho_{xy}(B)$ appears around the quantum Hall value of $\rho_{xy}=h/2e^2$. The fits look reliable till around $V_g=0$ [Fig.~\ref{fig:e-h_Sum}(a)], where the obtained set is:  $n=1.53 \times 10^{15}$~m$^{-2}$, $\mu_e=4.42$~m$^2$/V$\cdot$s, $p=6.48 \times 10^{15}$~m$^{-2}$, $\mu_h=0.47$~m$^2$/V$\cdot$s. As seen from these and further estimations the changes in MR with increasing $V_g$ are mainly connected here with the drop in the hole density while the changes in $n$ are not substantial and limited within the range of $n=(1.5-2) \times 10^{15}$~m$^{-2}$ for the whole set of negative and positive $V_g$. The weakness of the electron density dependence on $V_g$ is well seen for large $V_g> \quad \sim3.5$~V where the weak field part of $\rho_{xy}(B)$ shows a monotonous linear behavior due to a single kind of carriers -- electrons only: $\rho_{xy}=B/en$, with the slope almost independent on $V_g$ (Fig.~\ref{fig:VgXX-XY}).

With a further increase of $V_g$ some factors of uncertainty appear in the overall MR. The first warning is that the QH features shift below the fundamental values of $\rho_{xy}=h/ie^2$, initially for the distorted feature at $i=2$ for $V_g\geqslant+1.5$~V [Figs.~\ref{fig:e-h_Sum}(a,b)], then for the plateau with $i=1$ for $V_g\geqslant+2.5$~V [Figs.~\ref{fig:e-h_Sum}(c,d)].
The development of this tendency leads to appearing additional inversion points in $\rho_{xy}(B)$: Fig.~\ref{fig:e-h_Sum}(c) for $V_g=+2$~V and, with the further lowering of  $\rho_{xy}(B)$, for $V_g=+2.5$~V [Fig.~\ref{fig:e-h_Sum}(d)] and $V_g=+3$~V (Fig.~\ref{fig:VgXX-XY}). The nature of these inversion points is certainly different from the discussed above so that using them as a fitting point becomes problematic. Another characteristic feature used in fitting -- a linear slope in  $\rho_{xy}(B)$ at high classic fields -- disappears at large $V_g$. Thus, description by the classical two-carrier model becomes doubtful in this range of $V_g$. The origin of these transformations may be in the admixture of additional electron states to the states of the hole magnetic levels as in speculations around Fig.~\ref{fig:Model_v-c}.

A guess to explain this complicated behavior in classical terms is that it is a manifestation of an additional kind of particles with lower mobility. Their nature may be connected with the complicated structure of the valence subband top where an electron-like curvature of $E(k_\parallel)$ exists in some range of wave vector values around zero between the lateral maxima. These particles should behave like heavy electrons as the $E(k_\parallel)$ curvature is lower here than in the conduction subband. The addition of low mobility electron component to conductivity will cause a modification of Equations~(\ref{eq:2r}) into Equations~(\ref{eq:3car}), obtained by the reversal of conductivity tensor into the resistivity tensor:

\begin{widetext}
\begin{subequations}
\label{eq:3car}
\begin{equation}
\rho_{xy}=\frac{B}{|e|}
\frac{(p-n_1-n_2)B^4+[p(b_1^2+b_2^2)-n_1(b_2^2+b_h^2)-n_2(b_1^2+b_h^2)]B^2+pb_1^2b_2^2-n_1b_2^2b_h^2-n_2b_1^2b_h^2}{(p-n_1-n_2)^2B^4+[(n_1b_2+n_2b_1)^2+c.p.1+2(n_1n_2b_h^2-n_2pb_1^2-n_1pb_2^2)]B^2+(n_1b_2b_h+c.p.2)^2},
\label{eq:3rxy}
\end{equation}
\begin{equation}
\rho_{xx}=\frac{1}{|e|}\frac{(n_1b_1+n_2b_2+pb_h)B^4+[n_1b_1(b_2^2+b_h^2)+n_2b_2(b_1^2+b_h^2)+pb_h(b_1^2+b_2^2)]B^2+b_1b_2b_h(n_1b_2b_h+n_2b_1b_h+pb_1b_2)}
{(p-n_1-n_2)^2B^4+[(n_1b_2+n_2b_1)^2+c.p.1+2(n_1n_2b_h^2-n_2pb_1^2-n_1pb_2^2)]B^2+(n_1b_2b_h+c.p.2)^2},
\label{eq:3rxx}
\end{equation}
\end{subequations}
\end{widetext}
with \begin{eqnarray*}
&&c.p.1\equiv (n_2b_h+pb_2)^2+(n_1b_h+pb_1)^2,\nonumber\\
&&c.p.2\equiv n_2b_1b_h+pb_1b_2,\nonumber\\
&&b_{1,2,h}\equiv 1/\mu_{1,2,h}.
\end{eqnarray*}

\begin{figure}[b]
\includegraphics[width=\columnwidth]{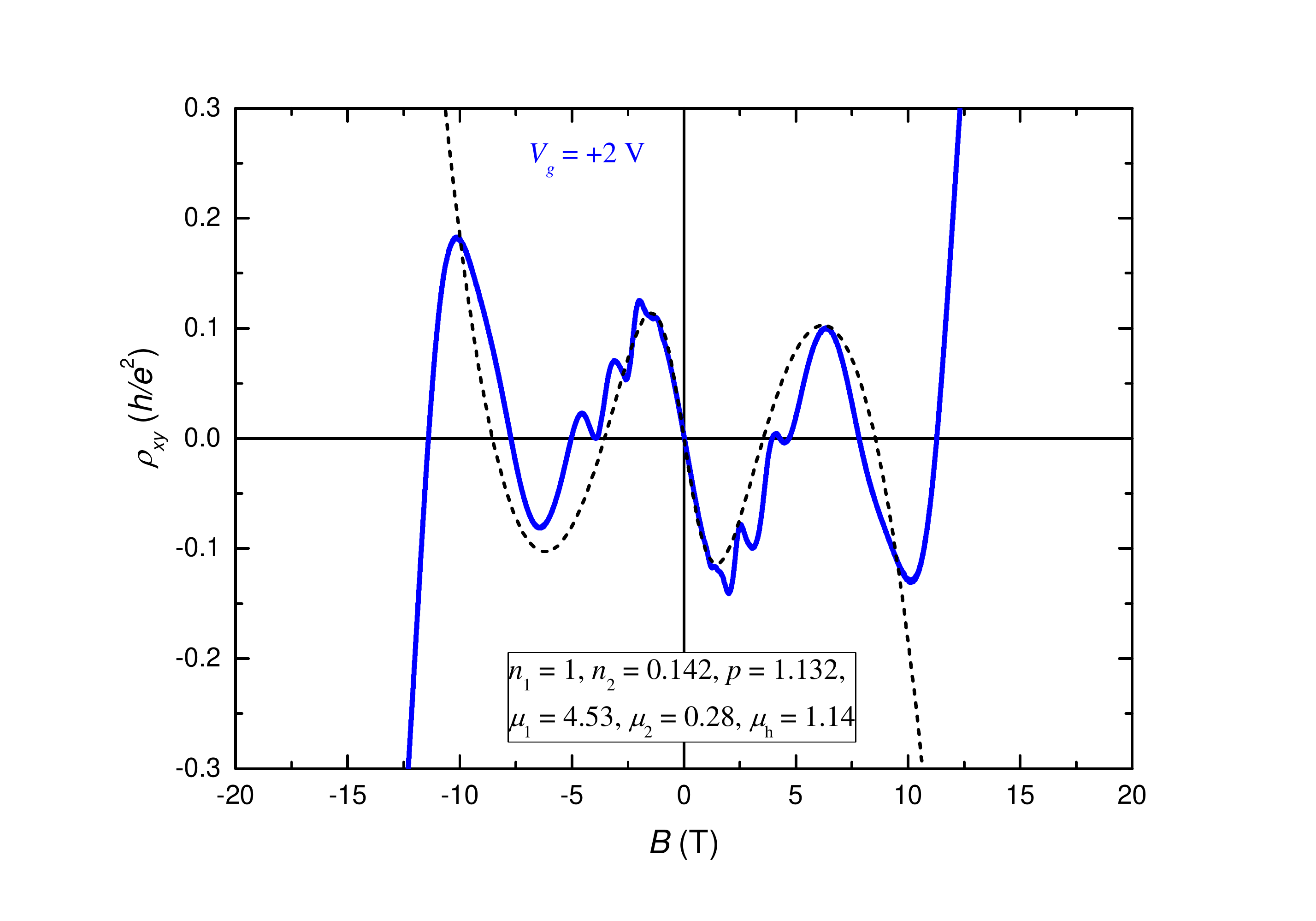}
\caption{\label{fig:eeh}  (color online). Three-carrier fit to $\rho_{xy}(B)$ for $V_g=+2$~V.}
\end{figure}

Using Egs.~(\ref{eq:3car}) it is indeed possible to obtain a second reversal of  $\rho_{xy}(B)$ (Fig.~\ref{fig:eeh}) for a realistic set of parameters: $n_1=1 \times 10^{15}$~m$^{-2}$, $\mu_1=4.53$~m$^2$/V$\cdot$s, $n_2=0.14 \times 10^{15}$~m$^{-2}$, $\mu_2=0.28$~m$^2$/V$\cdot$s,  $p=1.13 \times 10^{15}$~m$^{-2}$, $\mu_h=1.14$~m$^2$/V$\cdot$s ($n_1$, $n_2$, $\mu_1$ and $\mu_2$ are the densities and the mobilities of two kinds of electrons). Of course, these values are less reliable than in case of a single inversion as here are 6 fitting parameters and the overall fitting is used within a range of weak fields. Thus this approach is rather an illustration of possible manifestation of heavy electrons. The next inversion of $\rho_{xy}(B)$ to positive values in Fig.~\ref{fig:eeh} cannot be modeled in the three-carrier fit. A second kind of heavy holes with a still larger mass is needed for this. Thus, the task will be transformed into a four-carrier model. The necessity to add extra holes is seen from that $\rho_{xy}(B)$ is positive at the highest fields thus indicating that the sum of the hole densities is larger than that for the electrons, contrary to the presented values obtained in the three carrier fit. A nature of the second kind of holes may be suggested as being due to an asymmetry of the lateral maxima in the valence subband thus causing different values of $dE/dk$ on its opposite slopes. Some asymmetry in the DQW potential profile will lead to that different parts of the valence subband top would participate in the magnetotransport in the two layers thus providing heavy electrons from one layer and the second type of holes from the other one. We did not perform a four-carrier fit because of complexities and unreliable values expected for the 8 variable fit, but qualitatively the situation seems feasible once the classical approach works.

\begin{figure*}
\includegraphics[width=160mm]{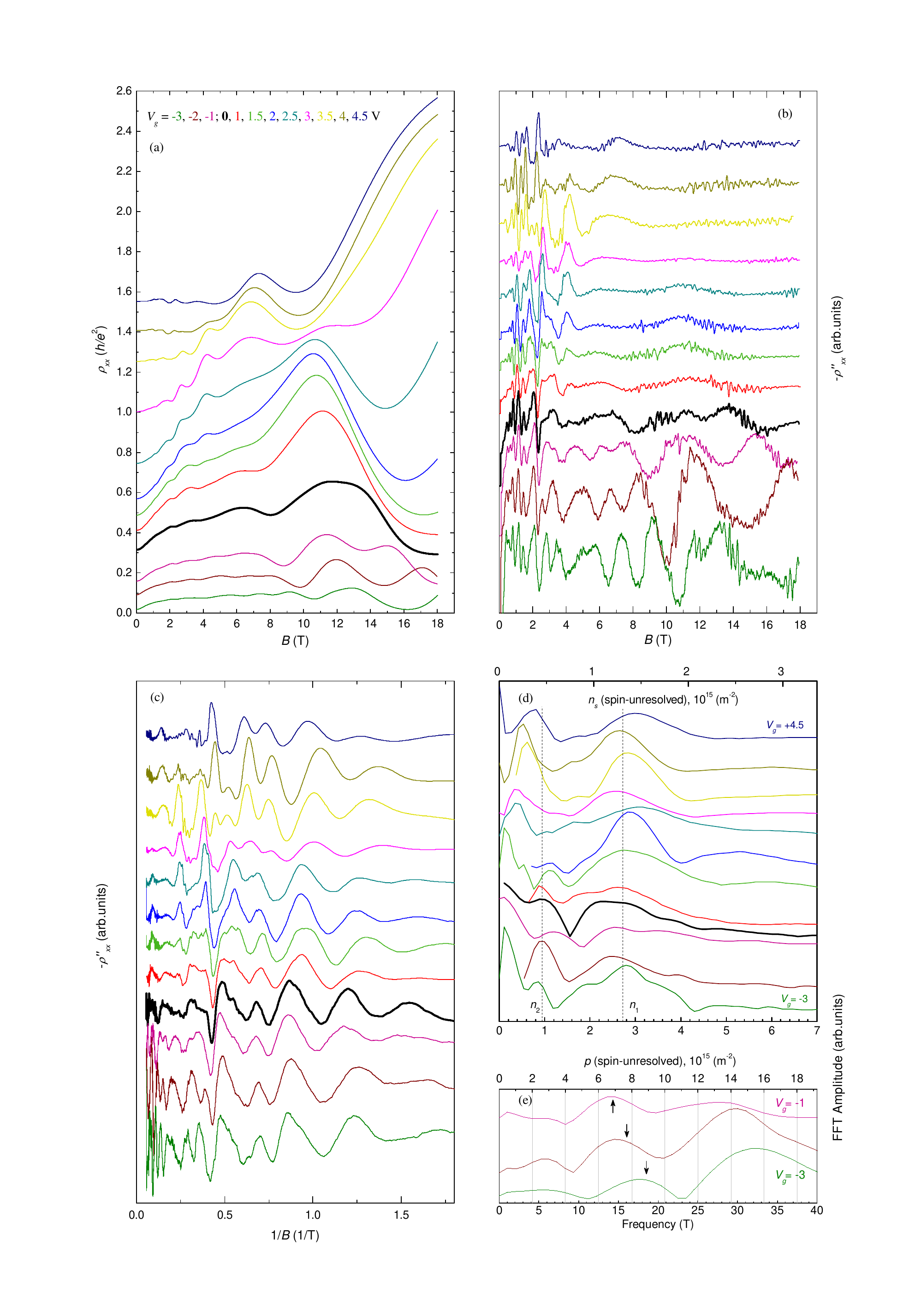}
\caption{\label{fig:Osc}  (color online). (a) $\rho_{xx}(B)$ of Sample 1 for different $V_g$, vertically shifted for clarity from the lowest $V_g=-3$~V at the bottom to the highest $V_g=+4.5$~V at the top; (b) their second derivatives $-d^2\rho_{xx}/dB^2$ versus magnetic field; (c) $-d^2\rho_{xx}/dB^2$ v.s. reciprocal field and (d,e) their FFT transforms. (d): The FFT results for the low-field windows corresponding to the low densities of electrons. Two dashed verticals correspond to some averaged positions of FFT peaks for $n_1$ and $n_2$. (e): The FFT results for the high-field windows corresponding to the high densities of holes. The arrows indicate $p$ values obtained from the classical two-carrier treatment. }
\end{figure*}

\subsection{\label{sec:level3b}Oscillations in $\rho_{xx}(B)$}

The uncertainty in obtaining parameters of free charge carriers from the classical magnetotransport at $V_g \gtrsim 0$ is a motivation to look for other ways to find them. Experimental curves for $\rho_{xx}(B)$ contain reach pictures of oscillations, although masked by the large and nonmonotonous backgrounds: Figs.~\ref{fig:VgXX-XY}, \ref{fig:SawS} and \ref{fig:Osc}~(a). We extract the oscillations by taking their second derivative $-d^2\rho_{xx}/dB^2(B)$, Fig.~\ref{fig:Osc}~(b), that considerably enhances fast oscillations in the weak field region. Oscillations demonstrate several periods in reciprocal magnetic fields, $-d^2\rho_{xx}/dB^2(1/B)$, Fig.~\ref{fig:Osc}~(c), that is expectable for several populated subbands. To do the analysis feasible, considering weak initial amplitudes and a limited number of oscillations, we performed separate Fast Fourier Transformations (FFT) in the regions of high fields, where the high density holes manifest themselves in the high field oscillations for the negative $V_g$, and within a low field window, where the oscillations should be due to the low density electrons: Fig.~\ref{fig:Osc}~(d,~e). The upper axis is added for the electron and hole densities related to the FFT frequency $f$ as $n,p=2fe/h$ for the unresolved spin splittings.

\begin{figure}[b]
\includegraphics[width=90mm]{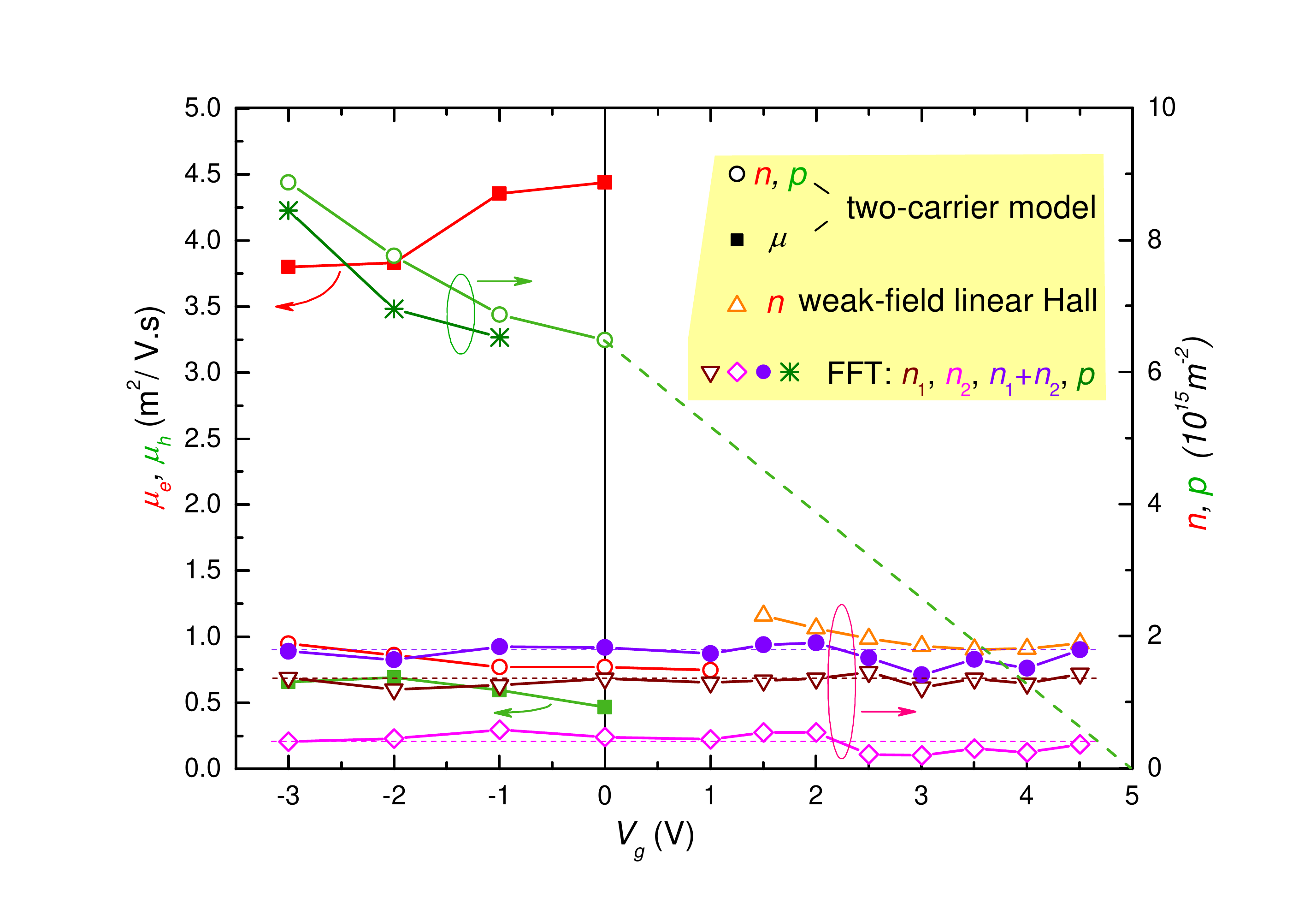}
\caption{\label{fig:npS}  (color online). The results on the electron (lines and symbols of the red tint) and hole (green tints) densities (circles, triangles, diamonds, asterisks) as well as mobilities (squares) obtained for Sample 1 in the two-carrier fits (empty circles for densities), weak field linear Hall (up-triangles) and FFT (down-triangles and diamonds for the two electron components and solid circles for their sum, asterisks for holes).}
\end{figure}

In the low frequency range the FFT curves reveal two main peaks: one around $n_1=1.5 \times 10^{15}$~m$^{-2}$ and the other at  $n_2\approx0.5 \times 10^{15}$~m$^{-2}$ as depicted by two dashed verticals in Fig.~\ref{fig:Osc}~(d). No directed movement of these peaks with $V_g$ is seen. In the high frequency window [Fig.~\ref{fig:Osc}~(e)] FFT peaks are revealed for the set of $V_g$ between $-1$~V and $-3$~V. One peak is close in its positions to the hole densities found from the two-carrier fits (indicated by arrows), if evaluations are performed for the spin unresolved case as $p=2fe/h$. The second peak seen in Fig.~\ref{fig:Osc}~(e) at approximately double frequencies should be connected with the spin resolved component of the same oscillations and yields the same $p$ values if estimated for the spin resolved case as  $p=fe/h$. The stable directed shift of the hole FFT amplitudes is seen with the changes of $V_g$ to negative values.

The results of all estimations for all kinds of carriers are collected in Fig.~\ref{fig:npS}: for two-carrier fits that yield summary densities for electrons and holes and their mobilities for $V_g\leqslant 0$; for simple weak-field classical linear $\rho_{xy}(B)$ yielding $n$ at maximum positive $V_g$ and for Fourier treatment of oscillations. When compared FFT results for electrons and the estimates for $n$ from the two-carrier fit
it should be considered that in a latter case the fit would not feel the possible distribution of electrons between two subbands if they have comparable mobilities in the subbands, while FFT feels the difference in their densities. Then the sum $n_1+n_2$ obtained from FFT should be compared with $n$ obtained from two-carrier fit. As seen from Fig.~\ref{fig:npS} the dots for $n_1+n_2$ are between the data for two-carrier fits and simple classical linear weak-field $\rho_{xy}(B)$ in a reasonable agreement. The general conclusion about electrons is that $n$ is almost independent on $V_g$. 

We have reliable data on $p$ only for $V_g\leqslant 0$ both for two-carrier fits and FFT analysis. However we can add one supplementary estimation for the positive $V_g$. The Hall MR $\rho_{xy}(B)$ in its evolution with increasing $V_g$ becomes fully negative at $V_g=3.5$~V: Fig.~\ref{fig:VgXX-XY}. It should happen when $n$ exceeds $p$. Thus, we can add this CNP to the function $p(V_g)$: $p=n$ at $V_g=3.5$~V. The dashed line in Fig.~\ref{fig:npS} depicts this rough prolongation of $p(V_g)$ to the positive $V_g$. The general conclusion on $p$: it pronouncedly decreases with $V_g$ from $p \approx 9 \times 10^{15}$~m$^{-2}$ at $V_g=-3$~V to zero at about $V_g=+5$~V.

\subsection{\label{sec:level3c}Analysis based on the energy band calculations}

\begin{figure}[b]
\includegraphics[width=90mm]{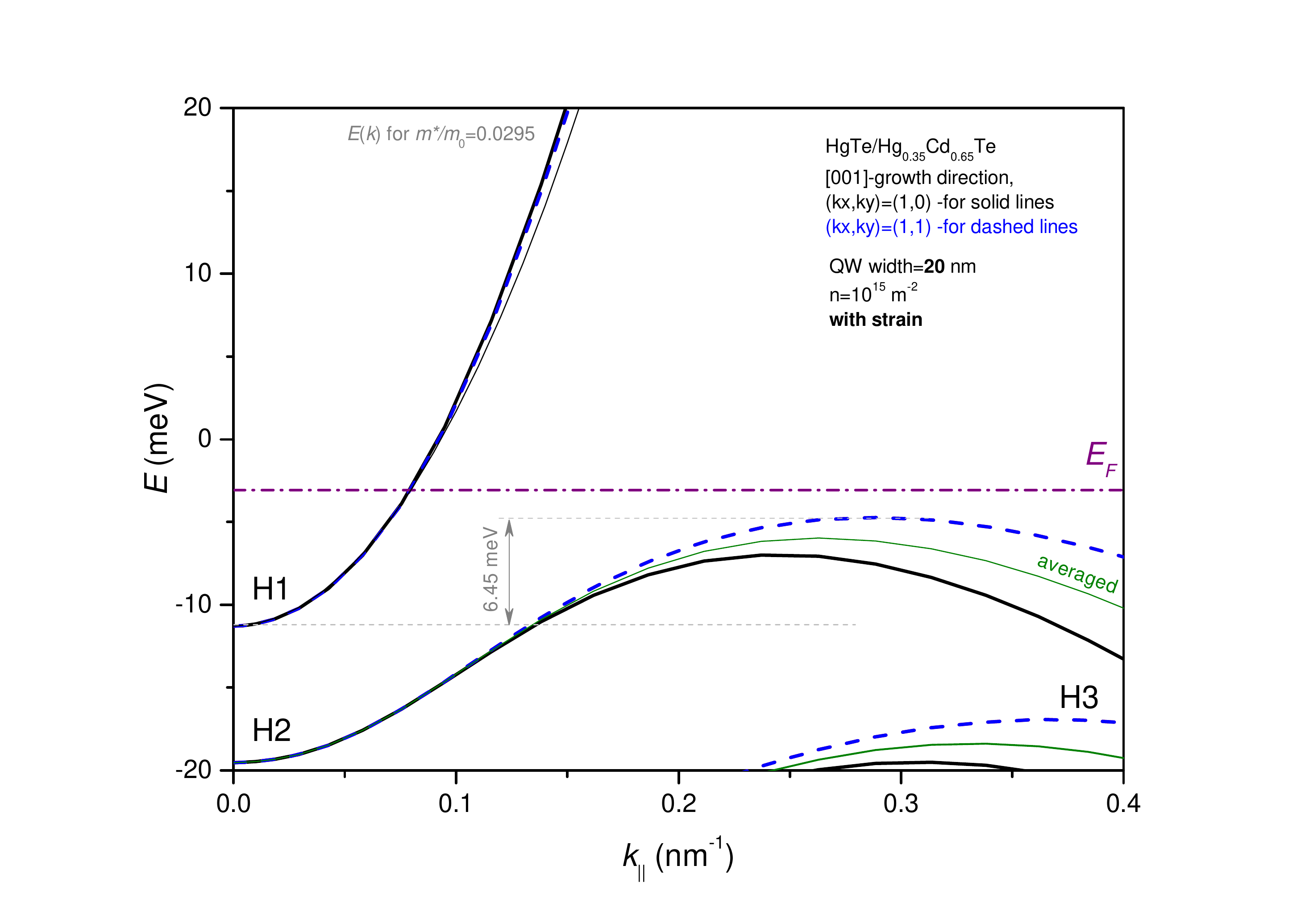}
\caption{\label{fig:E(k)SQW}  (color online). The calculated band structure for a single quantum well with parameters for the layers of which our DQW is built. Bold solid lines are for $(k_x,k_y)=(1,0)$, dashed lines -- for $(k_x,k_y)=(1,1)$. Thin solid curves -- for the averaged structures of the bands. Dot-dashed horizontal -- an example of the Fermi level position for $n=1 \times 10^{15}$~m$^{-2}$.}
\end{figure}

The energy band structure of a single fully strained QW, of which our DQW is built, was calculated in an envelope function approach  within the framework of 8-band $\mathbi{k}\cdot\mathbi{p}$ theory self-consistently with the Poisson equation for the charge distribution.\cite{ENovik} We used the simple [001] orientation for calculations in spite of [013] growth direction in our structures hoping that the $18^\circ$ deviation between these directions would not influence considerably the results of our analysis, which is mainly of qualitative character. The resultant band structure is presented in Fig.~\ref{fig:E(k)SQW}. 

In the valence subband, $E_v(k_{||})$ substantially differs for the (1,0) and (1,1) directions in the $(k_x,k_y)$ plain (bold solid and dashed curves in Fig.~\ref{fig:E(k)SQW} respectively), so that the constant energy contours are warped. We ignored this warping in our rough estimations of the carrier densities and used the averaged curve for the valence subband dispersion (thin green curves in Fig.~\ref{fig:E(k)SQW}).

A substantial overlap about 6.5~meV of the valence and conduction subbands is obtained when the strain is considered in calculations (while the overlap is absent at all without strain). But this overlap would not be felt experimentally in a single QW at electron densities $n > \quad \sim 1 \times 10^{15}$~m$^{-2}$ since the Fermi level is above the overlap region. Thus, attempts to use this picture without any modifications for interpretation of our experiments on DQWs would not be successful as we have  $n > 1.5 \times 10^{15}$~m$^{-2}$.

\begin{figure}[b]
\includegraphics[width=90mm]{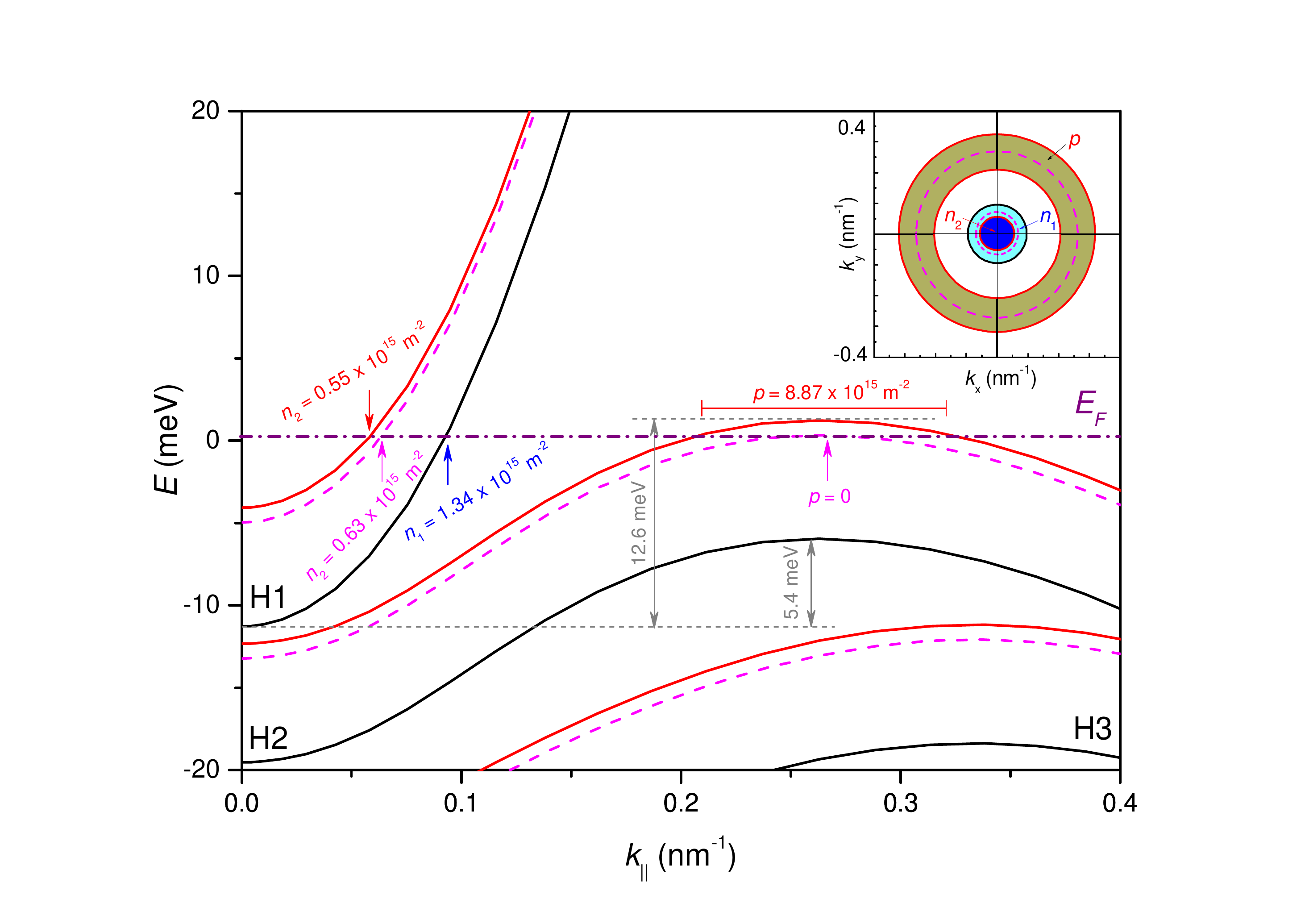}
\caption{\label{fig:DQW_E(k)ins}  (color online). The band structure of a double quantum well obtained by superposition of two similar pictures for the single QW (see Fig.~\ref{fig:E(k)SQW}) with an energy shift for the upper layer (upper curves of the red tint) with respect to the lower one (black lower curves). Energy of the upper layer is sensitive to $V_g$ while $E_F$ is fixed in the buffer structure below DQW. Solid red lines are for $V_g=-3$~V and the pink dashed lines -- for the moment when becomes $p=0$ (expected at $V_g=+5$~V). Notice that the simultaneous change in $n$ is tiny. Inset: The corresponding Fermi contours in the $(k_x,k_y)$ plain (for the averaged energy in the valence subband so that the warping is ignored). Notice a huge difference in the filled areas of the valence and conduction subbands.}
\end{figure}

Calculations of the energy spectra for HgTe/CdHgTe DQW\cite{Michetti-Novik2012-2014} show that the tunneling gap between the $\Gamma_8$ states is vanishingly small. In this case the DQW energy spectrum may be obtained by a simple superposition of pictures calculated for single QWs composing the DQW, with a relative shift in energy if some electric field presents across the structure. Here remarkable is that we observe the QH plateau with $i=1$ in the DQW even for zero gate voltage. For zero tunneling gap the $i=1$ QH state is impossible in a DQW with symmetric potential profile. In this case the  $i=1$ QH plateaus in each of the composing equal single QWs would yield the resulting Hall resistivity of the whole structure twice as smaller as for each layer switched in parallel thus corresponding to $i=2$ in the measurements on the DQW sample. In other words, the equal QWs would have equal sets of magnetic levels and combining these layers into a DQW structure of symmetric potential profile would simply lead to the doubling of all levels for zero tunneling gap. Thus the only possibility for appearing the gaps with odd filling factors in this case is an energy shift between the two layers. The existence of such a shift at $V_g=0$ may be due to asymmetric distribution of fixed charges: ionized dopants and uncontrolled impurities, surface charges, etc., or due to some differences in the shape of QWs.

\subsubsection{\label{sec:level3c-1}Free charges distributions between subbands}

We can reconstruct the energy spectrum of our DQW and its evolution with $V_g$ using the data on the free charges in the subbands. First, we take an energy spectrum calculated for a single QW and determine where the Fermi level should be drawn in this picture to get the experimentally obtained $p$. E.g., for  $p=8.87 \times 10^{15}$~m$^{-2}$ measured at $V_g=-3$~V, we calculate the area $S_k$ occupied by holes in the two-dimensional $k$-space using the relation $p=S_k/2\pi^2$ for the spin-degenerated states and adjust the $E_F$ position with respect to the lateral maxima of the valence subband to get this area of the ring cut from it by $E_F$: see the solid red upper valence subband in Fig.~\ref{fig:DQW_E(k)ins} with the obtained $E_F$ position and the corresponding ring in the inset. The derived $E_F$ cuts a part of the conduction subband of this layer (see again the solid red upper curve but for the conduction subband), and a similar calculation yields $n_2=0.55 \times 10^{15}$~m$^{-2}$. This value is less than the measured $n=1.89 \times 10^{15}$~m$^{-2}$. The missing part of electrons is in the conduction subband of the other (lower) layer of the DQW: $n_1=1.34 \times 10^{15}$~m$^{-2}$. To fulfill this condition the energy spectrum of the lower layer is superposed onto that of the upper layer (black lower curves in Fig.~\ref{fig:DQW_E(k)ins}) with a shift down on the energy of 7.2~meV that yields the area cut by $E_F$ from the conduction subband of the lower layer for the needed value of $n_1$. Thus the value of the energy shift between the layers is determined. This superposition results in an increased overlap between the conduction and valence subbands from 5.4~meV within a single layer to 12.6~meV between the layers at the expense of inclusion of the interlayer overlap. The overlapped states of the two layers contribute to the common conductivity of the DQW, also the interlayer Coulomb interactions between them may result in a formation of the interlayer electron-hole correlated states and other processes.

Variation of $V_g$ causes a shift of the upper layer energy with respect to the lower layer energy while the latter is insensitive to $V_g$ as the lower layer is screened from the upper gate by the free charges in the upper layer. The upper layer energy also shifts with respect to $E_F$, which is thus fixed in the lower layer and the part of the structure below.  As we obtained experimentally, the increase of $V_g$ to positive values causes a drop of $p$ from a substantial value to zero while $n$ remains almost unchanged. The found DQW spectrum (Fig.~\ref{fig:DQW_E(k)ins}) allows to explain this behavior. (i) The main part of the total electron density $n$ is in the lower layer conduction subband ($n_1$), which is insensitive to $V_g$. (ii) A very large density of states corresponds to the holes around the lateral maxima. These holes occupy a ring (a warped ring, to be more exact) in the 2D $k$-space with an area an order of magnitude larger than for circle areas corresponding to electrons: Fig.~\ref{fig:DQW_E(k)ins}(inset). A large effective mass corresponds to these holes. That is why a small decrease in the valence subband position with respect to $E_F$ causes a large drop in $p$. So, for the drop from $p=8.87 \times 10^{15}$~m$^{-2}$ (at $V_g=-3$~V) to zero (expected at $V_g=+5$~V) only a lowering of the upper layer energy on 0.9~meV is needed: see the dashed pink curves in Fig.~\ref{fig:DQW_E(k)ins}. This shift causes only a small increase of the electron density in the conduction subband of the upper layer, from $n_2=0.55 \times 10^{15}$~m$^{-2}$ to $0.63 \times 10^{15}$~m$^{-2}$, that amounts only about $4\%$ of the total increase in $n$, on the background of large $n_1$. This is also illustrated in the inset to Fig.~\ref{fig:DQW_E(k)ins} for the same shift: while the occupied valence subband ring shrinks into a pink dashed circle line with the lowering of the upper layer energy, the conduction circle for $n_2$ bordered with the red solid circle line expands only a little to the pink dashed circle contour whereas the black circle contour for $n_1$ remains unchanged.

This interpretation confirms the existence of two components in the electron conductivity seen in the Fourier spectra of oscillations where two FFT peaks for $n_1$ and $n_2$ were obtained: Fig.~\ref{fig:Osc}(d). The DQW spectrum presented in Fig.~\ref{fig:DQW_E(k)ins} explains why the $i=+1$ QH state exists in DQW: the valence subband of the lower layer is shifted much below $E_F$ in the whole range of $V_g$. Thus, it is always empty of holes and don't contribute to the hole component of the QHE.

Why did we choose a scheme with the energy of the upper layer shifted above the energy of the lower layer? It is due to the observed dynamics of magnetotransport pictures with $V_g$. In the opposite case (swap the colors in Fig.~\ref{fig:DQW_E(k)ins} between the red tint and the black) $p$ will be independent of $V_g$ as it will be in the valence subband of the lower layer and all the changes with increasing $V_g$ should be at the expense of large shifts down of the upper layer conduction subband energy, i.e., for the large increase in $n_2$ and of the total electron density $n$, that is in drastic contrast with all our experimental data.

\subsubsection{\label{sec:level3c-2}Peculiarities in the QH range of fields}

\begin{figure}[t]
\includegraphics[width=90mm]{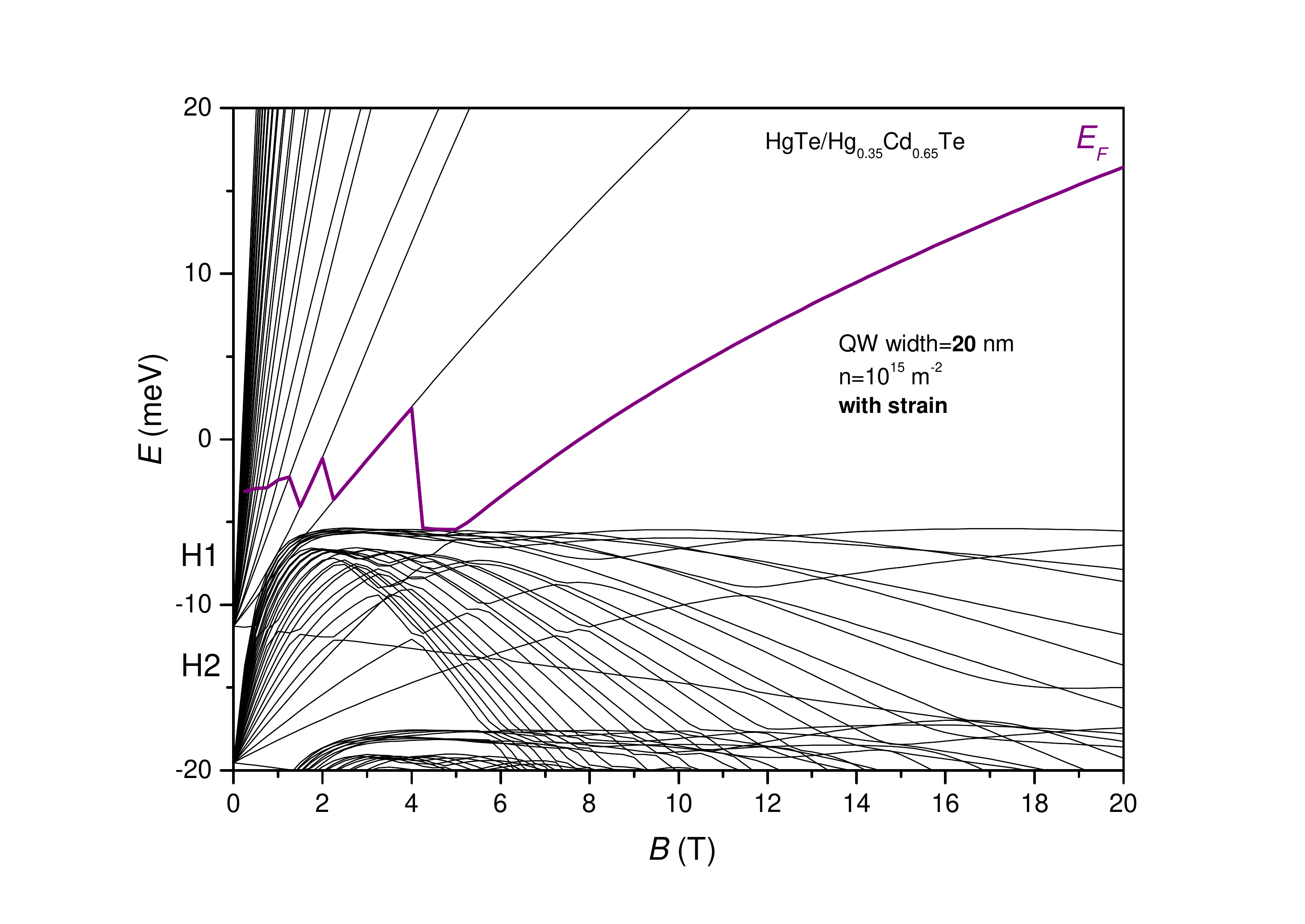}
\caption{\label{fig:LLs_SQW}  (color online). Picture of magnetic levels calculated for the energy spectrum of  Fig.~\ref{fig:E(k)SQW} with an example of the schematic Fermi level dynamics for $n=1 \times 10^{15}$~m$^{-2}$.}
\end{figure}

The scheme for computing $E(k_\parallel)$ was developed further for calculations of magnetic levels.\cite{ENovik} The magnetic level picture for the spectrum presented in Fig.~\ref{fig:E(k)SQW} is demonstrated in Fig.~\ref{fig:LLs_SQW}. Here an example of the Fermi level dynamics for $n=1 \times 10^{15}$~m$^{-2}$ is shown for the simplest case of $\delta$-shaped density of states in the levels. It is noteworthy that while $E_F$ does not touch the valence subband in $E(k_\parallel)$ for this $n$ (Fig.~\ref{fig:E(k)SQW}) it nevertheless enters the set of valence magnetic levels between 4 and 5~T in Fig.~\ref{fig:LLs_SQW}. This may be connected with  the anomalous nature of the lowest electron magnetic level. It is a so called zero-mode level, it starts from the valence subband at $B=0$ but monotonously moves up with field, as it should be for an electron level, while all the other valence subband levels, also starting from the valence subband at $B=0$ and moving initially like electron levels, turn down later at a certain field as they reach the energy of the lateral maximum thus behaving further like the hole levels. A nonmonotonous  behavior of the main set of the valence subband levels is well understood from a quasiclassical view for the valence subband with the lateral maxima, but the anomalous behavior of the zero-mode level does not have a quasiclassical interpretation. We believe that this zero-mode level yields a negative contribution to the Hall voltage thus proving its electron character. Another zero-mode level, but of the hole nature, is seen in Fig.~\ref{fig:LLs_SQW}: it starts at $B=0$ from the conduction subband but monotonously moves down as it should be for the hole level. Being the hole level it should not contribute to the conductivity as it should be empty of holes when $E_F$ is above it. In fact the movement of $E_F$ is more complicated within the levels of finite width especially within the field ranges where the electron and hole levels overlap, according to the modeled picture of levels in Fig.~\ref{fig:Model_v-c}.

\begin{figure*}
\includegraphics[width=160mm]{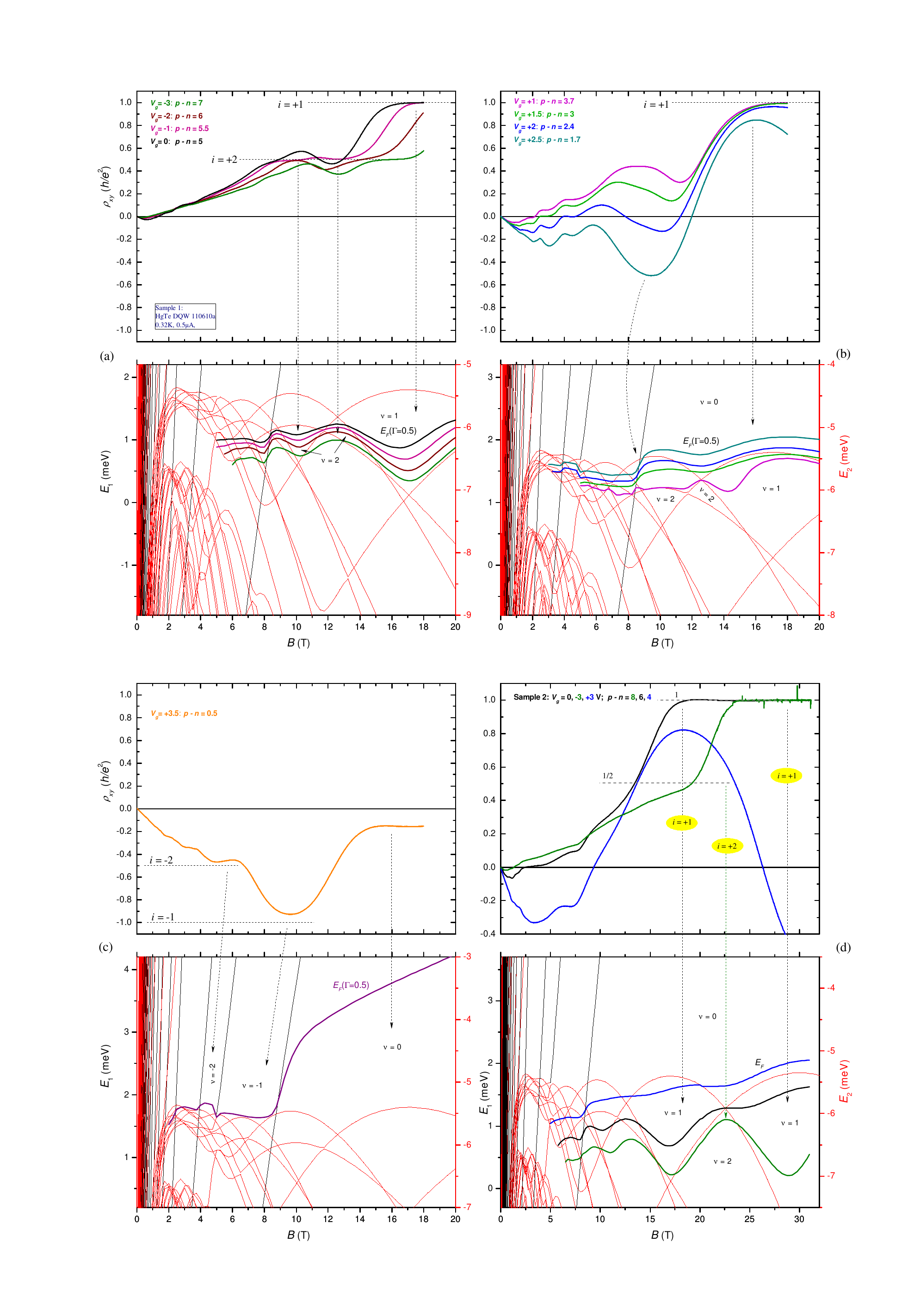}
\caption{\label{fig:S1S2EF}  (color online). Quantum Hall features in $\rho_{xy}(B)$ of Sample 1 (a,b,c) and Sample 2 (d) interpreted on the base of calculated DQW magnetic levels and obtained free charge densities. Red (black) magnetic levels are for the upper (lower) layer as in Fig.~\ref{fig:DQW_E(k)ins}. (a) for $V_g\leqslant 0$; (b) for $V_g>0$ in the range of multiple inversions; (c) in the vicinity of the charge neutrality point. $E_F$ is calculated the same way as the one presented in Fig.~\ref{fig:Model_v-c}.}
\end{figure*}

The magnetic level pattern for the DQW is obtained similarly as it was done for the DQW energy spectrum in Fig.~\ref{fig:DQW_E(k)ins}: the chart of levels calculated for the single QW in Fig.~\ref{fig:LLs_SQW} is superimposed onto the same chart with the energy shift on the already known value of 7.2~meV. We used the so obtained diagram of DQW levels for interpretation of the observed features in the QH range of fields. Of course, this interpretation is far from to be exact, but we hope at least to understand some of the main trends. The Fermi level dynamics within the obtained plot of the DQW magnetic levels was calculated for the finite width Lorentzian shaped density of states in the levels similarly to that in Fig.~\ref{fig:Model_v-c}. In these schematic calculations the levels in the valence subband were assumed to be of the hole nature within the whole range of fields (in spite of their electron-like behavior at low fields), i.e., their contribution to $\rho_{xy}(B)$ was programmed as positive, with exception of the zero-mode level, which was programmed as electron level as well as nonanomalous levels in the conduction subband. More detailed consideration of the nature of states in the levels is left for the future. The results of our calculations are presented in Fig.~\ref{fig:S1S2EF} for different free charge densities $(p-n)$ corresponding to the values of $V_g$ in experiment and compared to the measured $\rho_{xy}(B)$.

The $i=1$ QH plateau observed at $V_g=0$ corresponds to the position of $E_F$ inside the triangle-shaped gap formed by the arc-shaped course of a level near the lateral maximum at the top and the positively and negatively sloped levels on its flanks: Fig.~\ref{fig:S1S2EF}(a). The field $B_1$ when $E_F$ passes through the middle of the gap in this triangle correlates with QH plateau on $\rho_{xy}(B)$. With decrease of $V_g<0$ the tendency is seen as the calculated $B_1$ values move to the higher fields in agreement with our experiment. The transition region between QH states for $i=1$ and $i=2$ corresponds to the left flank of the triangle. The movement of the $(i=1) - (i=2)$ PPT to higher fields with decreased $V_g$ correlates with the shifts of the point where $E_F$ intersects the left flank. 

The gap for filling factor $\nu=2$ to the left from the $\nu=1$ triangular gap is formed by two negatively sloped and one positively sloped levels. It differs from the $\nu=1$ gap in shape and is much narrower. $E_F$ passes through this gap in the vicinity of three points where the levels cross. Existence of groups of closely spaced levels was observed to cause a similar distortion of the QH plateau in the shape of a dip (e.g., in a traditional DQW\cite{Y2008,Y2009}). A possible cause for appearance of such a dip on the QH plateau may be connected with an interplay between current carrying strips at the sample edges,\cite{Siddiki2014} and close positions of the neighboring levels may promote the effect.

With increase of $V_g>0$ the hole density decreases and the calculated Fermi level shifts upwards, finally penetrating into a gap above the lateral valence subband maxima and below the lowest electron level: Fig.~\ref{fig:S1S2EF}(b,c). In the joint picture of the hole and electron levels, population of the hole levels is described by the positive filling factors: $\nu=+1,+2,...$ and population of the electron levels -- by the negative ones: $\nu=-1,-2,...$. Then the gap above the valence subband lateral maxima is formally designated by $\nu=0$. It is characterized by an interference of the hole and electron states resulting in a drop of $\rho_{xy}(B)$ to zero in the vicinity of CNP. Simultaneously $\rho_{xx}(B)$ diverges: Figs.~\ref{fig:VgXX-XY} and \ref{fig:SawS}, as it was observed in a single HgTe QW at much smaller fields in the local\cite{Raichev-2012} and nonlocal\cite{Gusev-2012} transport. Possible physics causing a sharp drop of $\rho_{xy}(B)$ to zero with simultaneous growth of $\rho_{xx}(B)$ may be connected with a formation of the interlayer electron-hole excitons\cite{OlshanetskyJETPL2013} causing specific ratios for scattering probabilities between different channels.\cite{Raichev-2012}

$E_F$ on passing with increased $V_g$ through the valence subband edge at intermediate fields goes repeatedly across the field regions where it is either below or above the hole magnetic levels: Fig.~\ref{fig:S1S2EF}(b,c). This is due to the oscillating profile of the valence subband top formed in quantized magnetic fields by nonmonotonous course of the hole levels with different numbers around the energy of lateral maximum. Because of that and the superimposed electron levels the balance of the states on $E_F$ having the hole or electron character is alternatively changed. The contribution of electron states is enhanced when $E_F$ goes along the electron level between 8 and 9~T. All this results in several inversions in the course of $\rho_{xy}(B)$.

Speculations developed on the modeled level picture presented in Fig.~\ref{fig:Model_v-c} that the movement of the hole QH features with $V_g$ is stopped when $E_F$ reaches a crossing point between the hole  and electron levels may be justified in the pictures of the calculated levels shown in Fig.~\ref{fig:S1S2EF}. Here the first stopping point is the crossing between two valence subband levels around $B=13$~T (the left corner of the actual $\nu=1$ triangular gap), which appears due to that the hole level behaves in an electron manner to the left from the field where it reaches the lateral maximum. This crossing point is responsible for the stop of moving in field of the $(i=1) - (i=2)$ PPT. Another actual crossing point also appears in the calculations (which is more in the spirit of the model in Fig.~\ref{fig:Model_v-c}): the crossing of the lowest electron level with the valence subband edge between $B=8$ and 9~T. This point correlates with the minimum formed in $\rho_{xy}(B)$ around 10~T, which sinks to negative values at almost constant fields manifesting an increase in contribution of the electron states to magnetotransport at $E_F$. 

Similar behavior of quantum magnetotransport is observed at higher fields in Sample~2 (Fig.~\ref{fig:S1S2EF}(d)). A distinct $i=+1$ QH plateau at $V_g\leqslant +2$~V (see also Fig.~\ref{fig:SawS}) correlates with location of $E_F$ inside the triangular gap for $\nu=+1$. The curve for $V_g=-3$~V shows an inflection point at $\rho_{xy}(B)\approx 1/2$ that correlates with the passage of $E_F$ within the $\nu=+2$ gap, but closely to its upper edge, that is why this QH feature is weakly resolved. The $i=+1$ plateau is rather wide for $V_g=0$. It corresponds to the passage of $E_F$ through the two adjacent triangular $\nu=+1$ gaps crossing the neck between them. Surprisingly this crossing is not reflected in the plateau shape. Probably there is some anticrossing within the neck that might smooth the expected feature. The $i=+2$ feature is rather blurred here probable because the $\nu=+2$ triangle gap is too narrow. $E_F$ goes out into the $\nu=0$ gap at high fields for $V_g=+3$~V in agreement with the observed turn of $\rho_{xy}(B)$ to negative values. The oscillatory behavior of the valence subband edge is clearly manifested here as $E_F$ passes the positive $\nu=1$ gap close to its round top between $B=13$~T and 21~T resulting in a tendency for $\rho_{xy}(B)$ to reach the $i=+1$ plateau and its further turn back to negative values at the weaker fields.

\section{Conclusions}

In summary, we have found that the inverted energy spectrum of the quasi-2D HgTe is substantially modified in the DQW structure built of these layers. The overlap of the conduction and valence subbands may be considerably increased in the DQW. Due to this fact the critical field for opening the gap in the energy  spectrum shifts to much higher fields with respect to the critical field in a single QW. Thus, the accompanying specific features in magnetotransport, such as multiple inversions in  $\rho_{xy}(B)$, zero filling factor state with a concomitant manifestation of its insulator character in  $\rho_{xx}(B)$, etc., are moved to the range of higher fields where the QH regime is well realized. The effects of interlayer screening of the gate potential, which is never exhausted in the HgTe DQW, makes it possible to shift the energy spectra of the layers with respect to each other thus making possible to regulate the overlap between the conduction and valence subbands by the gate voltage $V_g$. A huge difference in the density of states is found between the valence subband lateral maxima and the conduction subband minimum leading to that only the hole density is sensitive to $V_g$ while the electron density remains almost constant.

\begin{acknowledgments}
Authors are grateful to E. Palm, T. Murphy, J. H. Park, and G. Jones for help with the experiment. The research was carried out within the state assignment of FASO of Russia (theme 'Spin' No.~01201463330), supported in part by RFBR (project No.~14-02-00151) and the RAS Ural division complex program (project No 15-9-2-21). National High Magnetic Field Laboratory is supported by NSF Cooperative Agreement No.~DMR-1157490 and the State of Florida.\end{acknowledgments}

\end{document}